\documentclass[epj,referee]{svjour}

\usepackage{amsmath}
\usepackage{amssymb}
\usepackage{graphics}

\begin{document}

\title{One-Component-Plasma: Going Beyond Debye-H{\"u}ckel}

\author{Andr{\'e} G.\ Moreira\thanks{e-mail: amoreira@mpikg-teltow.mpg.de}
and Roland R.\ Netz}

\institute{Max-Planck-Institut f{\"u}r Kolloid- und Grenzfl{\"a}chenforschung,
Kantstr. 55, 14513 Teltow, Germany}

\date{Received: date / Revised version: date}

\abstract{
Using field-theoretic methods, we calculate the internal
energy for the One-Component Plasma (OCP). We go beyond the 
recent calculation
by Brilliantov [N.\ Brilliantov, Contrib.\ Plasma Phys.\ {\bfseries 38}, 489 (1998)] 
by including non-Gaussian terms. 
We show that, for the whole range of the plasma parameter $\Gamma$,
the effect of the higher-order terms is small 
and that the final result is not
improved relative to the Gaussian theory when compared to simulations.
\PACS{{52.25.Kn}{Thermodynamics of plasmas}  \and
      {61.20.Gy}{Theory and models of liquid structure} \and
      {05.20.-y}{Statistical mechanics}} 
}

\maketitle

\section{Introduction}

In its simplest form, the One-Component Plasma (OCP) consists of a collection
of $n$ equally charged point-particles immersed in a neutralizing background
that assures global charge neutrality of the system. 
The OCP is important in several areas of physics as a starting 
point from where concepts or more refined theories are derived.
For instance, 
in astrophysics both OCP and its quantum-mechanical counterpart 
(the electron gas or ``jellium'') are used in the description of degenerate
stellar matter (interior of white dwarfs and outer layer of neutron
stars) and the interior of massive planets like Jupiter\cite{Ishimaru}.
In condensed matter physics, jellium is often
used as a reference state when calculating the electronic structure of solids.
When generalized to a Two-Component Plasma (or Restricted Primitive Model, if
hard-core interactions are taken into account), 
it can describe electrolytes and electrostatically stabilized colloidal solutions. 
For reviews see \cite{Ishimaru,Baus-Hansen}.

Different analytical techniques 
were employed in order to understand the OCP.
These were, in most cases, based 
on integral equations (as for instance in \cite{Gillan,Lado,macGowan})
or modified Mayer expansions\cite{Abe,Cohen}, i.e., 
low density expansions that used infinite
resummation of diagrams that accounted for the long range
character of the Coulomb interaction.
Comparison of the theoretical results with experimental data
is usually not possible. It is here where simulations 
(or ``computer experiments''\cite{Baus-Hansen}) play a 
particularly important role, by
providing a test ground where the suitability and
range of validity of the different approaches can be checked.
In general, the simulations use Monte Carlo 
technique\cite{Brush,Hansen-Weiss,Slattery1,Slattery2} and
yield quantities like the internal energy or the pair distribution
function $g(r)$.

In a recent paper\cite{Brilliantov1}, a field-theoretic
approach was used to treat the OCP. The introduction
of a cut-off at small wavelength (large-$k$) related to the
mean distance between particles
led to a good agreement, for all values of the plasma parameter $\Gamma$,
between the calculated internal energy 
and simulation results. The field-theoretic action used in \cite{Brilliantov1}
neglects terms other
than the Gaussian ones, i.e., it goes up to second order
in the fluctuating field. Here we extend this by including
more terms in the action and calculating consistently, using the same cut-off,
their contribution to the internal energy. As our main result, we show that
the higher order terms do not affect significantly the
results obtained with the Gaussian theory derived in \cite{Brilliantov1}.

\section{The Field-Theoretic Model}

Let us assume a classical system 
where $n$ positively charged particles are immersed in
a neutralizing negatively charged background.
The partition function of this system is
\begin{multline}
  \label{part1}
  Z = \frac{1}{n!} \biggl[ \prod_{j=1}^{n} \int 
  \frac{{\mathrm{d}}\mathbf{r}_j}{\lambda^3} \biggr]
  \exp \biggl\{ 
  E_{\mathrm{self}} \\
  - \frac{1}{2} \int {\mathrm{d}} \mathbf{r}
  {\mathrm{d}} \mathbf{r}^{\prime} \, \hat{\rho}_c( \mathbf{r}) \,
  v( \mathbf{r}- \mathbf{r}^{\prime}) \, \hat{\rho}_c( \mathbf{r}^{\prime})
  \biggr\}
\end{multline}
where $\lambda$ is an arbitrary constant,
$v( \mathbf{r})=\ell_B / r$ is the bare Coulomb operator and
$\ell_B \equiv e^2 / 4 \, \pi \, \epsilon \, K_B \, T$  is the Bjerrum
length (the length at which two elementary charges have an interaction
energy equal to the thermal energy). The
charge density $\hat{\rho}_c( \mathbf{r})$ is defined as
\begin{equation}
  \label{c_dens_op}
  \hat{\rho}_c( \mathbf{r}) = - \rho_-  + q \, \sum_{j=1}^{n} 
  \delta ( \mathbf{r}- \mathbf{r}_j),
\end{equation}
where $q$ is the valency of the particles and $\rho_-$ is the
uniform density of the neutralizing background. Global charge neutrality
implies that $\rho_- = q \, n / V$, where $V$ is the volume of the system. 
The Coulomb self-interaction of the particles is given by
\begin{equation}
  \label{Eself}
  E_{\mathrm{self}} = \frac{q^2 \, n}{2} \, v(0) =
  \frac{q^2 \, n}{2} \, \int \frac{{\mathrm{d}} \mathbf{k}}{(2 \, \pi )^3}
  \, \frac{4 \, \pi \, \ell_B}{k^2}.
\end{equation}
This corresponds to an infinite shift in the chemical potential, which is unimportant
for the thermodynamics of the system. We keep this term for reasons that will
become clear later (cf.\ Eq.~\ref{f_DH} below).

We can apply the Hubbard-Stratonovich transformation and
obtain a partition function that depends on a fluctuating
field $\phi$. This follows closely what has been done in 
\cite{Netz-Orland}, and so we state here the final expression
\begin{multline}
  \label{part3}
    Z^{\mathrm{ex}} \equiv  Z \, e^{-S}
    = \int \frac{D\phi}{Z_0}
    \exp \biggl\{ E_{\mathrm{self}} \\ 
    - \frac{1}{2} \int {\mathrm{d}} \mathbf{r} \, 
    {\mathrm{d}} \mathbf{r}^{\prime} \, \phi( \mathbf{r}) \,
    v_{\mathrm{DH}}^{-1}( \mathbf{r}- \mathbf{r}^{\prime}) \, 
    \phi( \mathbf{r}^{\prime}) + W[\phi] \biggr\},
\end{multline}
where $S=-n \ln(c \, \lambda^3)$ is the ideal entropy of the
particles ($c = n / V$) and 
\begin{equation}
  \label{Z0}
  Z_0 = \int D\phi
  \exp \biggl\{- \frac{1}{2} \int {\mathrm{d}} \mathbf{r} \, 
  {\mathrm{d}} \mathbf{r}^{\prime} \, \phi( \mathbf{r}) \,
  v^{-1}( \mathbf{r}- \mathbf{r}^{\prime}) \, 
  \phi( \mathbf{r}^{\prime}) \biggr\}.
\end{equation}
We defined $Z^{\mathrm{ex}}$ as the ``excess partition function'', 
the part of $Z$ that accounts only for the interactions 
between the particles. The propagator $v^{-1}$ in (\ref{Z0}) 
is the inverse of the bare Coulomb operator and $v_{\mathrm{DH}}^{-1}$ 
in (\ref{part3}) is given by
\begin{equation}
  \label{DH}
  v_{\mathrm{DH}}^{-1}( \mathbf{r}) = - 
  \frac{\nabla^2 \delta( \mathbf{r})}{4 \, \pi \, \ell_B} + 
  q^2 \, c \, \delta( \mathbf{r}).
\end{equation}
It is easy to show that this propagator is 
the inverse of the Debye-H{\"u}ckel operator
$v_{\mathrm{DH}}( \mathbf{r})= \ell_B \, e^{\kappa \, r} /r$,
where $\kappa^{-1}$ is the screening length given by 
$\kappa^2 = 4 \, \pi \, \ell_B \, q^2 \, c$. The
$W[\phi]$ is an infinite series in $\phi$ that contains
only non-Gaussian terms. Up to eighth order it reads
\begin{equation}
  \label{W}
  \begin{split}
    W[\phi] =& \frac{{\mathsf i} \, I_3 \, V}{3!} \, \overline{\phi^3}+
    \frac{I_4 \, V}{4!} \,\biggl( \overline{\phi^4} - 
    3 \, \overline{\phi^2}^2 \biggr) \\ 
    &- \frac{{\mathsf i} \, I_5 \, V}{5!} \, \biggl(
    \overline{\phi^5} - 10 \, \overline{\phi^2} \, \overline{\phi^3} \biggr)\\
    &- \frac{I_6 \, V}{6!} \, \biggl(
    \overline{\phi^6} - 15 \, \overline{\phi^4} \, \overline{\phi^2} -
    10 \, \overline{\phi^3}^2 + 30 \, \overline{\phi^2}^3 \biggr) \\
    &+ \frac{{\mathsf i} \, I_7 \, V}{7!} \, \biggl(
    \overline{\phi^7} - 21 \, \overline{\phi^2} \, \overline{\phi^5} -
    70 \, \overline{\phi^3} \, \overline{\phi^4} + 210 \, \overline{\phi^2}^2 \,
    \overline{\phi^3} \biggr) \\
    &+ \frac{I_8 \, V}{8!} \, \biggl(
    \overline{\phi^8} - 28 \, \overline{\phi^6} \, \overline{\phi^2} -
    56 \, \overline{\phi^5} \, \overline{\phi^3} - 35 \, \overline{\phi^4}^2 \\
    &+
    420 \, \overline{\phi^4} \, \overline{\phi^2}^2 + 560 \,
    \overline{\phi^2} \, \overline{\phi^3}^2 - 630 \overline{\phi^2}^4 \biggr)
  \end{split}
\end{equation}
with $I_m \equiv q^m \, c$ and ${\mathsf i}^2=-1$;
to simplify the notation, we use
$\overline{\phi^n}=\int \mathrm{d} \mathbf{r} \, \phi^n / V$.

If we define
\begin{equation}
  \label{Z_DH}
  Z_{\mathrm{DH}} = \int D\phi \,
  \exp \biggl\{ 
  - \frac{1}{2} \int {\mathrm{d}} \mathbf{r} \, 
  {\mathrm{d}} \mathbf{r}^{\prime} \, \phi( \mathbf{r}) \,
  v_{\mathrm{DH}}^{-1}( \mathbf{r}- \mathbf{r}^{\prime}) \, 
  \phi( \mathbf{r}^{\prime}) \biggr\},
\end{equation}
then the excess free energy of the OCP is given by
\begin{equation}
  \label{free_E}
  \frac{F^{\mathrm{ex}}}{K_B \, T}= - \log Z^{\mathrm{ex}} = - E_{\mathrm{self}}
  - \log \biggl( \frac{Z_{\mathrm{DH}}}{Z_0} 
  \biggr) - 
  \log \bigl\langle e^{W[\phi]}
  \bigr\rangle.
\end{equation}
The angular brackets correspond to a Gaussian average
where the the inverse Debye-H{\"u}ckel operator (\ref{DH}) is used as 
propagator. The term
\begin{equation}
  \label{f_DH}
  \begin{split}
    \frac{F^{\mathrm{ex}}_{\mathrm{DH}}}{K_B \, T} &\equiv
    - E_{\mathrm{self}}
    - \log \biggl( \frac{Z_{\mathrm{DH}}}{Z_0} \biggr)  \\
    &= - \frac{V}{2 \, \pi^2} \int\limits_{0}^{\infty} \mathrm{d}k \, k^2
    \biggl[
    \frac{\kappa^2}{2 \, k^2} - \frac{1}{2} \,
    \log \biggl( 1+\frac{\kappa^2}{k^2} \biggr)
    \biggr]
  \end{split}
\end{equation}
is the Debye-H{\"u}ckel contribution to the free energy. Notice that
$E_{\mathrm{self}}$ automatically regularizes this integral
in the ultra-violet, allowing its evaluation without the need of a small
wave-length (large-$k$) cut-off. 

Using (\ref{f_DH}) and neglecting
the term $\log \bigl\langle e^{W[\phi]} \bigr\rangle$ in 
(\ref{free_E}), we get the excess free energy per
particle
\begin{equation}
  \label{f_ll}
  f^{\mathrm{ex}} \equiv 
  \frac{F^{\mathrm{ex}}}{n \, K_B \, T} = - \frac{1}{\sqrt{3}} \,
  \Gamma^{3/2}
\end{equation}
where $\Gamma \equiv q^2 \, \ell_B \bigl(4 \, \pi \, c / 3  \bigr)^{1/3}$
is the (dimensionless) plasma parameter. From the excess 
free energy, we can get the internal energy per particle
\begin{equation}
  \label{u_DH}
  u \equiv
  \frac{U}{n \, K_B \, T} = \Gamma \, 
  \frac{\partial f^{\mathrm{ex}}}{\partial \Gamma}=
  - \frac{\sqrt{3}}{2} \, \Gamma^{3/2}.
\end{equation}
This is the resulting $u$ for
what we call from now on the ``Gaussian theory without cut-off''.

Eq.~(\ref{f_ll}) is the well known Debye-H{\"u}ckel limiting law, which
is asymptotically exact for vanishing $\Gamma$. 
At $\Gamma \sim O(1)$ the expression (\ref{u_DH})
already yields poor results when compared to simulations
(cf.\ Fig.~1b).
At large $\Gamma$ (cf.\ Fig.~1a) this inadequacy is particularly clear:
fits to simulation data 
show a linear behavior\cite{Slattery2,Young} 
in the internal energy, and not a 3/2 power law.

Brilliantov\cite{Brilliantov1} calculated the Gaussian theory
as depicted above but introduced a modification, namely a 
large-$k$ cut-off. This is justified with ideas that
follow the Debye theory for the specific heat in solids, stating
a direct relation between the number of allowed $\mathbf{k}$ modes
in the system and the number of degrees of freedom $3 \, n$.
The allowed wave vectors would be approximately inside a spheres of radius 
$k_o = \bigl( 9 \, c \, \pi^2 \bigr)^{1/3}$, which is used to 
substitute the $\infty$  in the integral in (\ref{f_DH}). 
The agreement between the internal energy obtained 
with this cut-off and the results from Monte Carlo 
simulations\cite{Slattery1,Slattery2}
are good (cf.\ Figs.\ 1 and 2). What we will show next is that the
inclusion of contributions up to eighth order in $\phi$ coming from the term
$\log \langle e^{W[\phi]} \rangle$ does not change significantly this result.

When the cut-off $k_o$ is used in (\ref{f_DH}), we get the
Debye-H{\"u}ckel excess free energy per particle 
\begin{multline}
  \label{f_DH_cut}
  f^{\mathrm{ex}}_{\mathrm{DH}} = - \frac{3}{2} \, 
  \bigl( b \, \Gamma \bigr)^{3/2} \, 
  \arctan \biggl(\frac{1}{\sqrt{b \, \Gamma}} \biggr) \\  
  - \frac{3}{4} \, \biggl( b \, \Gamma - 
  \log \bigl( 1 + b \, \Gamma \bigr) \biggr)
\end{multline}
where $\Gamma$ is the plasma parameter and 
$b \equiv \bigl( 2 / \pi^2 \bigr)^{1/3} \, 2 / 3 $ (keeping the notation 
used in \cite{Brilliantov1}). Notice that in the limit 
$\Gamma \rightarrow 0$ this expression reduces to (\ref{f_ll}),
as it should. This is the excess free energy used
in \cite{Brilliantov1} to calculate the internal energy.

In order to go beyond the Debye-H{\"u}ckel level, we do the cumulant
expansion
\begin{equation}
  \label{cumul}
  \log \langle e^{W[\phi]} \rangle =
  \langle W[\phi] \rangle + \frac{1}{2} \, \biggl( \langle W^2[\phi] \rangle -
  \langle W[\phi] \rangle^2 \biggr) + \cdots
\end{equation}
Using (\ref{W}) and going up to eighth order in $\phi$, we obtain
\begin{multline}
  \label{deltaF}
  - \frac{\log \langle e^{W[\phi]} \rangle}{n} =
  \frac{\pi}{108} \,
  \biggl[ \chi_3 + \frac{3}{2} \, \langle \phi^2 \rangle^2 \,
  \chi_1 \biggr] \\
  + \frac{\pi}{144} \, \langle \phi^2 \rangle^3 \,
  \chi_1 - \frac{\pi}{432} \, \chi_4 
\end{multline}
where
\begin{equation}
  \label{phi_2}
  \begin{split}
    \langle \phi^2 \rangle &= \frac{\ell_B}{2 \, \pi^2} 
    \int\limits_0^{k_o} \mathrm{d}k \, \frac{k^2}{k^2 + \kappa^2} \\
    &= \frac{9 \, b \, \Gamma}{2} \, \Biggl[
    1 - \sqrt{b \, \Gamma} \, \arctan \biggl( \frac{1}{\sqrt{b \, \Gamma}}
    \biggr) \Biggr]
  \end{split}
\end{equation}
and
\begin{equation}
  \label{chi_n}
  \begin{split}
    \chi_m &\equiv 4 \, \pi \, \int\limits_a^{\infty}
    \mathrm{d}r \,
    \langle \phi(0) \, \phi(r) \rangle^m \\
    &= \frac{4}{\pi^2} \biggl( \frac{9 \pi}{4} \biggr)^m \!
    m^{m-3} \bigl( b \, \Gamma \bigr)^{3 \, (m-1)/2} \,
    \overline{\Gamma} \bigl( 3-m, m \pi \sqrt{b \, \Gamma} \bigr).
  \end{split}
\end{equation}
$\overline{\Gamma}(m,x)$ is the incomplete gamma function\cite{gammafunc}
and $a = \pi / \bigl(9 \, c \, \pi^2 \bigr)^{1/3}$ is a
small distance cut-off. Ideally, the integrals $\chi_m$ should be performed
in $k$-space with the momentum cut-off given by $k_o$.
However, for $m \geq 3$, the Fourier transformed integrals cannot be solved in a 
closed form; on the other hand, the integrals, when
written in real space are not difficult to calculate exactly, provided
a small distance cut-off.
Since a large-$k$ cut-off corresponds in real space roughly to a 
small distance cut-off, we introduced as an approximation
the small-$r$ limit $a$ such that $k_o = \pi / a$ 

Putting (\ref{f_DH_cut}) and (\ref{deltaF}) into 
(\ref{free_E}), we finally obtain the expression
for the excess free energy of the OCP with contributions
up to eighth order in $\phi$. The internal energy follows
as in (\ref{u_DH}). 

\section{Results and Discussion}

In Figs.~1a and 1b we show $u$ obtained
from simulations\cite{Slattery1,Slattery2} (black circles), from
the Gaussian theory {\itshape with}\cite{Brilliantov1} 
and {\itshape without} (Eq.~\ref{u_DH}) cut-off 
(respectively dashed line and dash-dotted line)
and  from the results obtained here with the higher order terms 
(full line).

In the strongly coupled regime ($\Gamma > 1$, Fig.~1a),
the inclusion of higher order terms does not affect
the results obtained with the Gaussian theory with cut-off; 
both results are indistinguishable on this scale. We also
calculate the relative error in $u$, defined as
\begin{equation}
  \label{rel_err}
  R_{\mathrm{err}} = 
  \frac{u - u_{s}}{|u_{s}|},
\end{equation}
where the subscript $s$ stands for simulation.
In Fig.~2a we show, for $\Gamma > 1$, $R_{\mathrm{err}}$ 
for the Gaussian theory with cut-off (black circles) and for 
$u$ with the higher order terms calculated here (white circles). 
The agreement between theory and simulation is good, with deviations
between $-2 \%$ and $2 \%$\cite{note1}.
Notice however that
for $\Gamma \lesssim 10$ the inclusion of the higher order terms 
make the results worse, relative to the Gaussian
theory with cut-off, when compared to simulation.

In the weakly coupled regime ($\Gamma \leq 1$, Fig.~1b),
this trend is confirmed. In Fig.~2b it is clear that 
the deviation between theory and
simulations are larger when the higher orders are included.
However, in both cases $R_{\mathrm{err}}$ increases
with decreasing $\Gamma$. This is not surprising though, since
$u$ goes to zero as $\Gamma$ decreases, making
the relative error very sensitive to small differences between
theory and simulation.

In Fig.~3 we assess the importance of the higher order corrections computed
here in comparison to the Gaussian theory with cut-off. 
The higher order terms of the excess free energy 
are given by (\ref{deltaF}): the first term in the rhs corresponds
to the sixth order in $\phi$ correction and two remaining ones to the eighth
order. We can then 
write down $u$ as a sum of three terms, viz.
\begin{equation}
  \label{uu}
  u = 
  u_{\mathrm{DH}} + \Delta u_6 + \Delta u_8
\end{equation}
where $u_{\mathrm{DH}}$ is the Gaussian contribution to the internal
energy coming from (\ref{f_DH_cut}). In Fig.~3 we
plot $\Delta u_6 / |u_{\mathrm{DH}}|$ and
$\Delta u_8 / |u_{\mathrm{DH}}|$. As we can see,
$\Delta u_6$ is approximately one order of magnitude smaller 
than $u_{\mathrm{DH}}$ and $\Delta u_8$ two
order of magnitude smaller than $u_{\mathrm{DH}}$. 
We expect that the inclusion of terms of order
higher than eight will not change significantly the picture given here.

In summary, we have calculated higher order contributions to the 
internal energy of the OCP. We have shown that (i) the effects 
of these higher order terms are small relative to the previously calculated 
Gaussian theory\cite{Brilliantov1} and (ii) they do not
improve the agreement between theory and simulation
This shows that the Gaussian theory with the cut-off $k_o$ introduced
such as to approximately include strong nearest-neighbor correlations in the
high-$\Gamma$ limit is very accurate for describing the OCP\cite{note2}. 

Our calculation consists of two major steps, viz., the 
expansion of the excess free energy (\ref{free_E}) in cumulants of
$\phi$ and the introduction of the cut-off $k_o$ suggested
in Ref.\ \cite{Brilliantov1}. In principle, the first step
can be improved systematically by including higher order
terms, while there is no clear recipe for improving the second one.
The results we obtained here are then a consequence of the 
approximate way of calculating the cut-off. Since the
precise value of $k_o$ is not uniquely determined, it may be treated as
a fit parameter. By fitting the theory to simulation at the high-$\Gamma$
limit we obtain $k_{\mathrm{fit}} \simeq 4.417 \, c^{1/3}$, which is close 
to the value $k_o = \bigl( 9 \, \pi^2 \, c \bigr)^{1/3} \simeq 4.462 \, c^{1/3}$ 
used by Brilliantov and also used by us.

\begin{acknowledgement}
  AGM acknowledges financial support from the Portuguese FCT through the grant
  Praxis XXI/BD/13347/97.
\end{acknowledgement}


\begin{thebibliography}{10}
\bibitem{Netz-Orland}
R.~R. Netz and H. Orland, Europhys.\ Lett. {\bf 45},  726  (1999). Notice that
in Eq.~14 of this reference (which is equivalent to Eq.~\ref{deltaF} here)
the factor $a^3 I_5 I_3 / 32$ should read $a^3 I_5 I_3 / 16$.
\bibitem{note1}
Although the deviations are small, we should notice that they are statistically
significant since the numerical error given by simulations in this range of
$\Gamma$ is below $\pm 0.02\%$\cite{Slattery2}.
\bibitem{note2}
As a side remark, we note that
the application of the cut-off $k_o$ to the Two-Component Plasma (TCP)
leads to unphysical results. At the Debye-H{\"u}ckel level,
the excess free energy per volume goes for small
$c \equiv (n_+ + n_-)/V$ as
$\mathtt{f}^{\mathrm{ex}} \sim 
- \bigl(4 \pi \ell_B q^2 c \bigr)^{3/2} / 12 \pi$ (the already
discussed Debey-H{\"u}ckel limiting law) and as
$\mathtt{f}^{\mathrm{ex}} \sim 
- (9/ \pi)^{1/3} \, \ell_B q^2 c^{4/3}$ for large $c$;
at this limit total free energy per volume is unbounded
from below and non-convex. At finite temperatures, 
the Maxwell construction leads to the coexistence between states with
a zero and an infinite density, which is clearly unphysical and
signals that this approach is not appropriate for the TCP.

This difference between the OCP and the TCP becomes more transparent
in a recent systematic low-density expansion of the free energy, where
the TCP expansion coefficients are shown to be dominated by ion-pair
formation\cite{Netz-low-c}
\end{thebibliography}

\newpage

\begin{figure}
  \caption{Internal energy ($u$) as function of the 
    plasma parameter $\Gamma$ in the range (a) $1 < \Gamma < 300$ and (b)
    $0 < \Gamma \leq 1$. The full line denotes the result obtained here,
    the dashed line the Gaussian theory with cut-off\cite{Brilliantov1},
    the dash-dotted line the $u$ obtained from
    the Debye-H{\"u}ckel limiting law (Eq. \ref{u_DH}).
    The points denote simulation results from Refs.\ \cite{Slattery1} 
    ($0 < \Gamma < 1$) and \cite{Slattery2} ($1 < \Gamma < 300$).}
\end{figure}

\begin{figure}
  \caption{The relative error $R_{\mathrm{err}}$ (Eq.\ \ref{rel_err}) as
    function of $\Gamma$ for (a) $1 < \Gamma < 300$ and (b) 
    $0.1 < \Gamma \leq 1$. The white circles represent the relative error for the
    $u$ obtained here with the higher order corrections; the
    black circles represent the relative error for the Gaussian theory
    with cut-off\cite{Brilliantov1}.}
\end{figure}

\begin{figure}
  \caption{Comparison of the sixth order and eighth order
    corrections with $u_{\mathrm{DH}}$. Notice that
    $\Delta u_6$ is roughly one order of magnitude
    smaller than $u_{\mathrm{DH}}$ and 
    $\Delta u_6$ is two orders smaller than $u_{\mathrm{DH}}$.}
\end{figure}

\end{document}